\documentclass[%
reprint,
amsmath,amssymb, aps,prl,
]{revtex4-1}

\usepackage{graphicx}
\usepackage{dcolumn}
\usepackage{bm}
\usepackage{hyperref}

\begin{document}

\title{Light polarization independent nuclear spin alignment in a quantum dot}

\author{E. A. Chekhovich$^{1,2}$, A. B. Krysa$^3$, M. S. Skolnick$^1$, A. I. Tartakovskii$^1$}

\affiliation{$^{1}$Department of Physics and Astronomy, University
of Sheffield, Sheffield S3 7RH, UK\\
$^{2}$Institute of Solid State Physics, 142432, Chernogolovka, Russia\\
$^{3}$Department of Electronic and Electrical Engineering,
University of Sheffield, Sheffield S1 3JD, UK }

\date{\today}

\begin{abstract}
We report optical pumping of neutral quantum dots leading to nuclear spin alignment with direction
insensitive to polarization and wavelength of light. Measurements of photoluminescence of  both ''dark'' and ''bright'' excitons in single dots reveal that nuclear spin pumping occurs via a virtual spin-flip transition between these states accompanied by photon emission. The sign of the nuclear spin polarization is determined by asymmetry in the exciton energy spectrum, rather than by the sign of the exciton spin polarization.
\end{abstract}

\maketitle


Control and understanding of the nuclear spin
environment in nano-structures is of great importance in achieving
robust coherence of spin-based qubits in the solid state.
Recently, optical pumping of nuclear spins in quantum dots (QDs) has
been demonstrated
\cite{GammonPRL,Eble,Skiba,GaAsDiffusion,Lai,ResInP,Latta,Xu1}.
Dynamic nuclear polarization (DNP) due to the electron-nuclear hyperfine interaction (HI) occurs when
non-equilibrium populations of electron spin states are created using resonant or quasi-resonant circularly polarized light of high intensity. Under such conditions, a direct correspondence
between the sign of circular polarization of the exciting light
and the direction of the nuclear spin alignment is observed
\cite{GammonPRL,Eble,Skiba,GaAsDiffusion,Lai,Tartakovskii}.
By contrast, suppression of the electron spin alignment under
above barrier non-resonant excitation results in
negligible nuclear spin polarization in a dot.

Here we report measurements on individual neutral InP/GaInP
quantum dots, which shed new light on the mechanisms of DNP in semiconductor nano-structures.
They reveal previously unobserved phenomena: at low pumping levels, {\it independent}
of light polarization and wavelength, optical pumping induces an effective nuclear field, which is
{\it always parallel} to the external field applied along the growth axis of the structure.
We show that at low power DNP in a neutral dot occurs via the second order recombination
of ''dark'' excitons accompanied by electron-nuclear spin flip-flop.
This is revealed in photoluminescence (PL) measurements where  optically inactive ''dark''
excitons are observed due to weak mixing with the ''bright'' states. Asymmetry in the energy splitting of the excitonic energy levels induced by electron-hole exchange interaction leads to a strong difference of
DNP rates induced by ''dark'' excitons with opposite spins \cite{Bracker2}. As a result, the direction of nuclear polarization is independent of the average electron spin polarization on the dot. It is instead  controlled by the direction of the external magnetic field experienced by the exciton.

\begin{figure}
\includegraphics{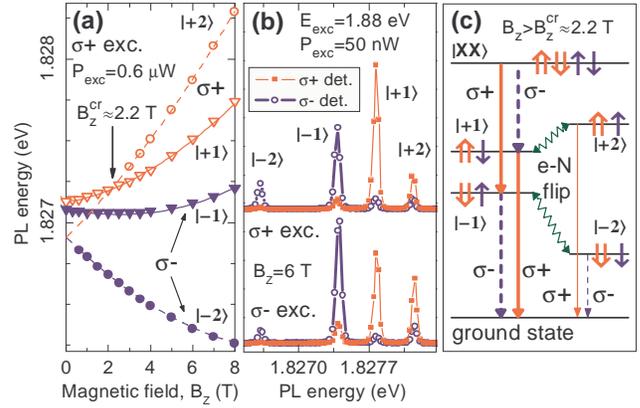}
\caption{\label{fig:SpecScheme} (a) Magnetic field dependence of
exciton PL energies (symbols), lines show fitting. (b) PL spectra
of the exciton in a neutral quantum dot detected in $\sigma^+$ (
$\blacksquare$) and $\sigma^-$ ($\bigcirc$) polarizations under
$\sigma^+$ (top) and $\sigma^-$ (bottom) excitation with photon
energy 1.88~eV at $B_z$=6.0~T. (c) Energy diagram of exciton and
biexciton states in high magnetic field
$B_z>B_z^{cr}\approx2.2$~T. Heavy holes $\Uparrow$($\Downarrow$)
and electrons $\uparrow$($\downarrow$) with spin parallel
(antiparallel) to external field form optically allowed
($\Uparrow\downarrow$, $\Downarrow\uparrow$) and ''dark''
($\Uparrow\uparrow$, $\Downarrow\downarrow$) excitons with total
spin $J_z=\pm1$ and $\pm2$ respectively. Thick arrows show
circularly polarized recombination of biexciton and ''bright''
excitons, thin arrows correspond to weak recombination of ''dark''
exciton states. Zigzag lines show electon-nuclear ($e$$-$$N$)
spin-flips induced by the hyperfine interaction.}
\end{figure}

By contrast in the regime of higher powers, where the exciton states of the dot are saturated, we find
a direct correspondence between the helicity of light and the direction of the nuclear spin alignment,
as was observed previously \cite{GammonPRL,Eble,Skiba,GaAsDiffusion,Lai,Tartakovskii}. In this regime DNP
is determined by optical orientation of electrons in excited states in the dot or in the wetting layer.

The experiments were performed on an undoped InP/GaInP sample without
electric gates. PL of neutral InP QDs was measured at $T=4.2$~K, in
external magnetic field $B_z$ up to 8~T normal to the sample
surface. QD PL at $\sim$1.84~eV was excited with a laser either
below ($E_{exc}$=1.88~eV) or above ($E_{exc}$=2.28~eV) the GaInP barrier
band-gap and analyzed with a 1~m double spectrometer and a CCD.

In a neutral dot electrons $\uparrow$($\downarrow$) and heavy
holes $\Uparrow$($\Downarrow$) with spin parallel (antiparallel)
to the growth axis $Oz$ can form either optically-forbidden
(''dark'') excitons $\left|\Uparrow\uparrow\right>$
($\left|\Downarrow\downarrow\right>$) with spin projection
$J_z=+2(-2)$, or ''bright'' excitons
$\left|\Uparrow\downarrow\right>$
($\left|\Downarrow\uparrow\right>$) with $J_z=+1(-1)$ optically
allowed in $\sigma^+$($\sigma^-$) polarization. The electron-hole
($e-h$) exchange interaction splits off ''dark'' excitons by the
energy $\delta_0$. An additional splitting $\delta_b$($\delta_d$)
of the bright(dark) exciton doublet is caused by the reduced
symmetry of the QD potential \cite{GammonPRL,Bayer}. QD axis
misorientation or symmetry reduction may lead to weak mixing of
''bright'' and ''dark'' states allowing observation of the latter
in PL \cite{Bayer,InPDyn}. Below we denote mixed exciton states
with spin projections $J_z\approx\pm2,\pm1$ as
$\left|\pm2\right>$, $\left|\pm1\right>$ distinguishing them from
''pure'' excitons
$\left|\Downarrow\uparrow\right>$, $\left|\Uparrow\downarrow\right>$, $\left|\Uparrow\uparrow\right>$
and $\left|\Downarrow\downarrow\right>$. PL energies of bright
(dark) excitons $E_{b(d)}$ measured as functions of $B_z$ in a
single neutral dot with $\delta_0\approx200$~$\mu$eV,
$\delta_{b(d)}\approx65(0)$~$\mu$eV are shown in Fig.
\ref{fig:SpecScheme} (a) with triangles (circles). Fitting of the
$E_{b(d)}(B_z)$ dependences [lines in Fig.\ref{fig:SpecScheme}(a)] allows electron (hole) g-factor $g_{e(h)}\approx+1.6(+2.7)$
to be extracted (see appendix in Ref. \cite{InPDyn} for more
details on QD characterization).

At high $B_z$ exceeding both the effective exchange field
$\delta_b/(\mu_B|g_h-g_e|)\approx$1~T and the field
$B_z^{cr}\approx$2.2~T, where $\left|+1\right>$ and
$\left|+2\right>$ states cross, PL originating from all four
exciton states can be resolved as shown in Fig.
\ref{fig:SpecScheme}~(b) where PL spectra detected in $\sigma^+$
($\sigma^-$) polarization at $B_z$=6.0~T under $\sigma^+$ and
$\sigma^-$ polarized excitation are shown by the thin (thick) lines.
The energy level structure of all exciton states as well as the biexciton
$\left|\Uparrow\Downarrow\uparrow\downarrow\right>$  at
$B_z>B_z^{cr}$ is shown in Fig. \ref{fig:SpecScheme}~(c) along
with possible optical transitions.

\begin{figure}
\includegraphics{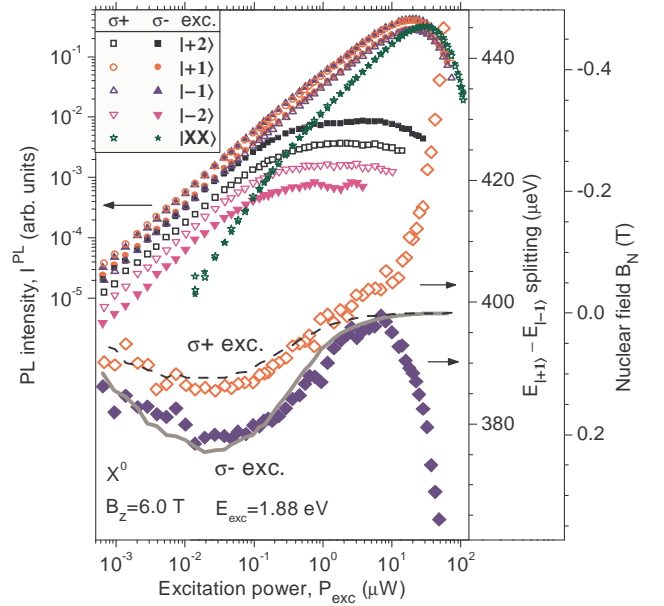}
\caption{\label{fig:PDep} Results of PL power dependence in a
neutral dot at $B_z$=6~T under $\sigma^+$ (open symbols) and
$\sigma^-$ (solid symbols) excitation with photon energy 1.88~eV.
PL intensities (left scale) for all four exciton states and for
the total intensity of biexciton emission are plotted. Spectral
splitting (right scale) of the two optically-allowed exciton
states ($E_{\left|+1\right>}-E_{\left|-1\right>}$) is shown with
diamonds. Lines show fitting obtained using the model presented in text. Additional scale on the right
shows nuclear field $B_N$ on the dot deduced from the splitting.}
\end{figure}

Fig. \ref{fig:PDep} shows PL intensities $I^{PL}$ (left scale)
of all four exciton transitions and the total biexciton intensity at
$B_z$=6.0~T as functions of optical power $P_{exc}$ of $\sigma^+$ and $\sigma^-$
polarized excitation with photon energy $E_{exc}=1.88$~eV. Due to their small
oscillator strengths saturation of dark excitons is observed at $P_{exc}\approx1$~$\mu$W, whereas bright
excitons and biexcitons saturate at much higher powers of
20~$\mu$W and 40~$\mu$W, respectively.

The net nuclear field $B_N$ along $Oz$ acting via the HI shifts
exciton states with electron spin $\uparrow$($\downarrow$) by
$+\mu_Bg_eB_N/2$ ($-\mu_Bg_eB_N/2$). This allows measurement of
$B_N$ at $B_z>B_z^{cr}$ from the spectral splitting $\Delta
E_{\left|+1\right>,\left|-1\right>}=E_{\left|+1\right>}-E_{\left|-1\right>}$
of the bright exciton doublet: $B_N\approx(\Delta
E_{\left|+1\right>,\left|-1\right>}^0-\Delta
E_{\left|+1\right>,\left|-1\right>})/\mu_Bg_e$. $\Delta
E_{\left|+1\right>,\left|-1\right>}^0$ is the bright exciton
splitting at $B_N=0$ that we deduce from pump-probe measurements
where the sample is kept in the dark for a sufficiently long time to allow
nuclear spins to relax ($>$200 s) \cite{InPDyn}. Fig. \ref{fig:PDep} shows the dependence
of $\Delta E_{\left|+1\right>,\left|-1\right>}$ and the corresponding
value of $B_N$  (right scale) on the intensity $P_{exc}$ of circularly polarized
light at $B_z$=6.0~T. Results for $\sigma^+$
($\sigma^-$) excitation plotted with open (solid) symbols show the
variation of $\Delta E_{\left|+1\right>,\left|-1\right>}$ with
$P_{exc}$, and demonstrate the occurrence of nuclear spin pumping.

As can be inferred from Fig.\ref{fig:PDep}, there are two
distinct regimes in the DNP. At low excitation power
($P_{exc}<10$~$\mu$W) positive $B_N>0$ is induced for \emph{both}
$\sigma^+$ and $\sigma^-$ polarized excitation.  By contrast, at
high powers ($P_{exc}>10$~$\mu$W) $\sigma^{+(-)}$ excitation
results in $B_N<0$ ($B_N>0$), similar to previous reports
\cite{GammonPRL,Eble,Skiba,GaAsDiffusion,Lai,Tartakovskii}. We
note however that such direct correspondence between the helicity
of the excitation and direction of nuclear spin polarization is
only observed in the regime of saturation and suppression of the
exciton and biexciton PL which is accompanied by the occurrence of
broad background emission in the PL spectrum. We thus conclude
that the build-up of $|B_N|$ observed in this regime cannot
originate from the HI with the ground state excitons in the QD.
The most likely source of DNP in this case is the HI of the nuclei
with the spin polarized electrons in the excited shells of the dot
or in the wetting layer. This is further confirmed in an
experiment with above GaInP barrier band-gap excitation: for a
laser with photon energy $E_{exc}=2.28$~eV high power DNP is
suppressed almost completely, as electrons lose their spin
polarization during energy relaxation. By contrast, the unusual
polarization-insensitive DNP observed in the low power regime
remains effective under this condition: independently of light
polarization it leads to $B_N\approx+0.2$~T, similar to the case of
$E_{exc}=1.88$~eV.

As we show in more detail in the model presented below, these observations allow us to
conclude that in the wide range of powers $P_{exc}<10$~$\mu$W DNP
is induced by virtual flips of the $\left|+2\right>$ dark
exciton into the intermediate $\left|+1\right>$ state followed by
recombination. This second order process dominates due to the asymmetry
of the exciton spectrum seen in Fig.1: the energy splitting between $\left|+2\right>$ and $\left|+1\right>$
is significantly smaller than for $\left|-2\right>$ and $\left|-1\right>$ (see detailed explanation below).
This results in $B_N>0$ for both $\sigma^+$ and $\sigma^-$ polarized excitation despite
considerable differences in the exciton populations with $\uparrow$ and $\downarrow$ electrons. The interplay between processes responsible for nuclear spin pumping and depolarization results in the previously unobserved strongly non-monotonic power-dependence of $B_N$ in Fig.2. The model presented below explains this behavior, and describes the underlying mechanisms quantitatively.

Transfer of spin polarization from the exciton to the nuclear system
requires $e-N$ spin-flip that transforms a bright exciton into dark
or vice versa [zigzag lines in Fig. \ref{fig:SpecScheme} (c)]. The
energy splittings between exciton states ($\sim$100~$\mu$eV)
significantly exceed the nuclear spin level separation
($\sim$0.1~$\mu$eV) thus raising the problem of energy
conservation \cite{Imamoglu}. The energy mismatch can be
compensated by the photon emitted during recombination
\cite{GammonPRL,ResInP,Bracker2} in a second-order process. For
example, if the QD contains a $\left|+2\right>$ ''dark'' exciton
it can make a virtual flip into the $\left|+1\right>$ state
increasing nuclear spin polarization by +1 [zigzag line Fig.
\ref{fig:SpecScheme} (c)]. At the second stage the ''bright''
$\left|+1\right>$ exciton recombines with emission of a $\sigma^+$
photon [thick solid arrow in Fig. \ref{fig:SpecScheme}(c)]. Such
processes which change nuclear spin polarization by $+1$($-1$) can
start from $\left|+2\right>$ or $\left|-1\right>$
($\left|+1\right>$ or $\left|-2\right>$) states [see Fig.
\ref{fig:SpecScheme}(c)]. For each initial state only one
intermediate state (with the same hole and opposite electron spin)
is possible. Spin flips are governed by the off-diagonal matrix
element
$V_{hf}=\left<\Uparrow\uparrow\right|H_{hf}\left|\Uparrow\downarrow\right>=\left<\Downarrow\uparrow\right|H_{hf}\left|\Downarrow\downarrow\right>$
of the hyperfine Hamiltonian $H_{hf}$ which is proportional to the
in-plane component of the fluctuating nuclear field
\cite{Merkulov} and can be estimated as $|V_{hf}|\approx
A_{hf}/(2\sqrt{N})\sim1$~$\mu$eV , where
$A_{hf}\approx$230~$\mu$eV \cite{Gotschy} is the electron spin
splitting corresponding to fully polarized nuclear spins in the
QD, and $N$$\sim$10$^4$$\div$10$^5$ is the number of nuclei in the
dot.

As PL peaks corresponding to $J_z\approx+2(-2)$ dark excitons are
mainly $\sigma^+$($\sigma^-$) polarized the wave-functions of
mixed states can be approximated as:
\begin{eqnarray}
\left|\pm2\right>=c_+\left|\Uparrow\uparrow\right>+s_+\left|\Uparrow\downarrow\right> \quad[c_-\left|\Downarrow\downarrow\right>+s_-\left|\Downarrow\uparrow\right>],\nonumber\\
\left|\pm1\right>=c_+\left|\Uparrow\downarrow\right>-s_+\left|\Uparrow\uparrow\right>
\quad[c_-\left|\Downarrow\uparrow\right>-s_-\left|\Downarrow\downarrow\right>],\label{eq:wf}
\end{eqnarray}
with mixing parameters $c_{\pm}$$=$$\cos\phi_{\pm}$,
$s_{\pm}$$=$$\sin\phi_{\pm}$ ($\phi_{\pm}$$\ll$1).

The nuclear spin pumping rate is proportional to (i) the magnitude
of the hyperfine mixing between the initial and intermediate
states which is proportional to $V_{hf}^2$ and inversely
proportional to the square of the energy splitting between them,
(ii) the recombination rate of the intermediate exciton state
$\cos^2\phi_{\pm}/\tau_r$ ($\sin^2\phi_{\pm}/\tau_r$) for
$\left|\pm1\right>$ ($\left|\pm2\right>$) and (iii) the probability
$p_i$ to find the dot in the initial state $i$. The total spin
pumping rate $w_{pump}$ is a sum of the individual rates for all
four possible initial states and can be calculated using
wavefunctions from Eq.\ref{eq:wf} as:
\begin{eqnarray}
w_{pump}=\frac{\cos2\phi_-V_{hf}^2}{\Delta E_{\left|-1\right>,\left|-2\right>}^2}\left(\frac{\sin^2\phi_-}{\tau_r}p_{\left|-1\right>}-\frac{\cos^2\phi_-}{\tau_r}p_{\left|-2\right>}\right)\nonumber\\
+\frac{\cos2\phi_+V_{hf}^2}{\Delta
E_{\left|+2\right>,\left|+1\right>}^2}\left(-\frac{\sin^2\phi_+}{\tau_r}p_{\left|+1\right>}+\frac{\cos^2\phi_+}{\tau_r}p_{\left|+2\right>}\right),\label{eq:wpump}
\end{eqnarray}
where $\Delta
E_{\left|+2\right>,\left|+1\right>}=(E_{\left|+2\right>}-E_{\left|+1\right>})$,
$\Delta
E_{\left|-1\right>,\left|-2\right>}=(E_{\left|-1\right>}-E_{\left|-2\right>})$.
The $p_i$ are proportional to $I_i^{PL}$, the PL intensities plotted in
Fig. \ref{fig:PDep}, and inversely proportional to the photon emission rate:
\begin{eqnarray}
p_{\left|\pm2\right>}=(I^{PL}_{\left|\pm2\right>}/I_0)\times\tau_r\times\sin^{-2}\phi_{\pm},\nonumber\\
p_{\left|\pm1\right>}=(I^{PL}_{\left|\pm1\right>}/I_0)\times\tau_r\times\cos^{-2}\phi_{\pm},\label{eq:popul}
\end{eqnarray}
where the normalization constant $I_0$ corresponds to the
detected PL intensity per one photon emitted by the dot.

For the splittings $\Delta E\lesssim400~\mu$eV and
$\tau_r\approx1$~ns we can estimate $w_{pump}\gtrsim10^4$~s$^{-1}$
from Eq.\ref{eq:wpump}. For an alternative scenario
involving phonon-mediated transitions between the ''dark'' and
''bright'' states the rate is below $10^1$~s$^{-1}$
\cite{hfPhononRelaxation} under similar conditions and can be neglected. It also
follows from Eq.\ref{eq:wpump} that the dark excitons play the
major role in DNP. This is not only a result of their increased
population at low $P_{exc}$ \cite{GammonPRL,KorenevSelfPol}, but
also a result of the high recombination rate
$\tau^{-1}_r\cos^2\phi_{\pm}$ of the intermediate ''bright''
states. According to Eq.\ref{eq:wpump} $\left|+2\right>$ and
$\left|-2\right>$ excitons contribute to $w_{pump}$ with opposite
signs. However, for small $P_{exc}=50$~nW at $B_z=6.0$~T we detect
$B_N\approx$+0.13 and +0.21~T corresponding to $\approx$5$\%$ and
10$\%$ nuclear spin polarization for $\sigma^+$ and $\sigma^-$
excitation respectively. This is observed despite the substantial
decrease (increase) of $\left|+2\right>$($\left|-2\right>$)
population when the laser polarization is changed from $\sigma^-$
to $\sigma^+$: $p_{\left|+2\right>}/p_{\left|-2\right>}$ is
$\approx$2.6 times smaller for $\sigma^+$ excitation [see
Fig.\ref{fig:SpecScheme}(b)]. This indicates that DNP induced by
the $\left|+2\right>$ exciton is more efficient, which is explained by
smaller energy splitting between the initial and intermediate
states for the $\left|+2\right>$ exciton: $\Delta
E_{\left|+2\right>,\left|+1\right>}^2\approx\Delta
E_{\left|-1\right>,\left|-2\right>}^2/4.5$ [see
Fig.\ref{fig:SpecScheme}(a)].

Nuclear spin depolarization induced by optical pumping can result
from fluctuations of the electron Knight field \cite{GammonPRL} or
electric field gradients interacting with nuclear quadrupole
moments \cite{PagetQRelaxation}. In our model we assume that
recombination or capture of an exciton can result in simultaneous
flip of a nucleus with probability $R$. In the low power regime
($P_{exc}<10$~$\mu$W) the capture/recombination rate is
proportional to the total PL intensity of all exciton and
biexciton transitions $I_{tot}^{PL}$. Thus the optically induced depolarization
rate can be approximated as:
\begin{eqnarray}
w_{dep}^{opt}=2R\times(I_{tot}^{PL}/I_0)\times(\mu_Bg_eB_N/A_{hf}),\label{eq:decay}
\end{eqnarray}
where $I_0$ is the same normalization constant as in
Eq.\ref{eq:popul}, and the factor of 2 arises from spin relaxation
during capture and recombination. The non-optically induced relaxation
rate (e. g. due to nuclear spin diffusion or coupling with phonons) can be approximated as
\begin{eqnarray}
w_{dep}^{dark}=\tau_{dark}^{-1}\times(S_{P}+S_{In})/2\times
N\times(\mu_Bg_eB_N/A_{hf}),\label{eq:decayDark}
\end{eqnarray}
where $S_{P}=1/2$, $S_{In}=9/2$, and $\tau_{dark}\approx200$~s is
the nuclear spin decay time in the dark at $B_z=6$~T \cite{InPDyn}.

In the steady-state condition
$w_{pump}=w_{dep}^{opt}+w_{dep}^{dark}$. Using
Eq.(\ref{eq:wpump},\ref{eq:popul},\ref{eq:decay},\ref{eq:decayDark})
we can explicitly express $B_N$ as a function of the PL intensities
$I^{PL}$ measured in experiment. We note that when
$I_{tot}^{PL}$ reaches its maximum value the dot is saturated, which means
that the sum of the probability to find the dot in the biexciton state
$p_{\left|XX\right>}=(I^{PL}_{\left|XX\right>}/I_0)\times\tau_r/2$
and all probabilities in Eq. \ref{eq:popul} is $\approx1$. This
gives an additional relation that allows to eliminate $I_0$ from
the final expression for $B_N$.

The calculated dependence $B_N(P_{exc})$ for $\sigma^{+(-)}$ polarized
excitation is shown with a dashed
(solid) line in Fig.\ref{fig:PDep}. The following magnitudes of the fitting parameters
were obtained: $R$$\approx$1$\cdot$10$^{-3}$, mixing factors
$\sin^2\phi_+$$\approx$1.4$\cdot$10$^{-2}$,
$\sin^2\phi_-$$\approx$1.0$\cdot$10$^{-2}$, and
N$\approx$1.1$\cdot$10$^{4}$. We also used $\tau_r$=1~ns.
The calculated curves are in good agreement with the experimental
results for $P_{exc}<10$~$\mu$W. In particular the non-monotonic
character in the low-power regime (with the $B_N$ maximum at
$P_{exc}\sim$50~nW and \emph{depolarization} towards
$P_{exc}\approx1$~nW and $P_{exc}\approx10$~$\mu$W) is well
reproduced and can now be explained as follows: At low powers
$P_{exc}\approx1$~nW, the rate $w_{pump}$ is very small due to low
exciton occupancy and non-optically induced relaxation with the
rate $w_{dep}^{dark}$ (see  Eq.\ref{eq:decayDark}) dominates.
$w_{dep}^{dark}$ remains constant with increasing $P_{exc}$,
whereas the exciton occupancy increases leading to higher
$w_{pump}$, and as a result a higher degree of nuclear polarization.
At $P_{exc}>50$ nW, the population of dark states gradually
saturates leading to saturation of $w_{pump}$. On the other hand,
the rate of optically induced depolarization (see
Eq.\ref{eq:decay}) increases linearly with power in a wide range
of $P_{exc}$, and as a result nuclear polarization decreases.

\begin{figure}
\includegraphics{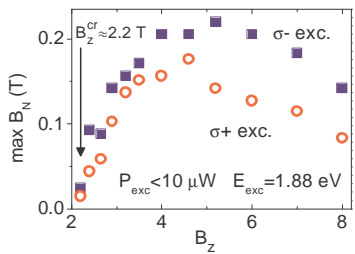}
\caption{\label{fig:BDep} Magnetic field dependence of the maximum
nuclear field $B_N$ induced at low power ($P_{exc}<10$~$\mu$W) of
$\sigma^+$ ($\bigcirc$) and $\sigma^-$ ( $\blacksquare$) polarized
excitation with photon energy $E_{exc}=1.88$~eV.}
\end{figure}

Finally we consider the dependence of the maximum $B_N$ induced in
the low power regime as a function of $B_z>B_z^{cr}$
shown in Fig.\ref{fig:BDep} for both helicities of below the GaInP barrier
excitation. Both curves have flat maxima at $B_z\approx5$~T. For higher fields,
$B_N$ decreases with $B_z$. This is consistent with the photon-assisted DNP interpretation
due to the vanishing difference between the splittings $\Delta
E_{\left|+2\right>,\left|+1\right>}^2\approx\Delta
E_{\left|-1\right>,\left|-2\right>}^2$.

It can also be seen from Fig.\ref{fig:BDep} that $B_N$ is
significantly reduced when $B_z$ approaches the crossing field
$B_z^{cr}\approx2.2~T$. This is a result of reduction of the
$w_{pump}$ rate (Eq. \ref{eq:wpump}) due to anticrossing
\cite{Bayer} of the $\left|+2\right>$ and $\left|+1\right>$
states, which originates from low-symmetry $e-h$ exchange
interaction. It leads to a minimum splitting between the two
corresponding PL lines $\delta_{exc}^e$$\approx20$~$\mu$eV at
$B_z=B_z^{cr}$. In experiment the anticrossing manifests itself as a
significant increase of the $\left|+2\right>$  PL intensity
as it gains oscillator strength from the $\left|+1\right>$ state. The
suppression of $w_{pump}$ takes place in the vicinity of $B_z^{cr}$ over a
field range equal to several times the value of $\delta_{exc}^e/(\mu_Bg_e)$.
Anticrossing acts in several ways: (i) by setting the lower limit
on $\Delta E_{\left|+2\right>,\left|+1\right>}$$=$$\delta_{exc}^e$
(see Eq.\ref{eq:wpump}), (ii) by reducing electron spin
projections due to the non-hyperfine mixing of $\left|+2\right>$
and $\left|+1\right>$ states (corresponding to $\phi_+$=$\pi/4$ in
Eq.\ref{eq:wf} at $B_z=B_z^{cr}$), and (iii) by reducing the
population of the $\left|+2\right>$ state due to increased
emission probability. We thus conclude that DNP insensitive to
light polarization should be enhanced for \emph{symmetric} quantum
dots where $\delta_{exc}^e\approx0$.

In conclusion, we have demonstrated optical pumping of nuclear
spins weakly dependent on the wavelength and polarization of the
low intensity light. The phenomena have been shown to originate from intrinsic asymmetry in
the energy spectrum of the excitonic energy levels. We note that this process may lead to \textit{self-polarization} - the optically induced spontaneous alignment of nuclear spins in
the limit of vanishing external field $B_z\approx0$ predicted for neutral quantum dots in Ref.
\cite{KorenevSelfPol}.

We thank K. V. Kavokin, A. J. Ramsay and M. N. Makhonin for
fruitful discussions. This work has been supported by  the EPSRC
Programme Grant EP/G601642/1 and the Royal Society. AIT is
grateful for support by an EPSRC ARF.


\end{document}